\documentclass[preprint,aps,floatfix,nofootinbib,a4paper,superscriptaddress]{revtex4-1}

\usepackage{amsmath, amssymb, amsfonts, amsthm, latexsym, epsfig, mathrsfs, xcolor, bbm, slashed}

\usepackage[inline]{enumitem}

\usepackage{setspace}
\usepackage[marginal, multiple]{footmisc}

\usepackage[utf8]{inputenc}

\usepackage[colorlinks, allcolors=blue!70!black, linktocpage]{hyperref}

\numberwithin{equation}{section}

\usepackage{cleveref}

\let\OLDthebibliography\thebibliography
\renewcommand\thebibliography[1]{%
	\setstretch{1.079} 
	\OLDthebibliography{#1}%
	\small %
	\setlength{\itemsep}{0.2\baselineskip} 
}

\let\OLDfootnote\footnote
\renewcommand\footnote[1]{%
	\setlength{\footnotesep}{0.75\baselineskip}%
	{\footnotesize \OLDfootnote{#1}}%
}

\setlist[enumerate]{noitemsep, label=(\arabic*), ref=(\arabic*)}

\renewcommand\thesection{\arabic{section}}
\renewcommand\thesubsection{\arabic{subsection}}

\makeatletter
\def\p@subsection{\thesection.}
\def\p@subsubsection{\thesection.\thesubsection.}
\makeatother

\theoremstyle{plain}

\newtheorem{prop}{Proposition}[section]

\theoremstyle{definition}

\theoremstyle{remark}
\newtheorem{remark}{Remark}[section]

\crefname{equation}{Eq.}{Eqs.}
\creflabelformat{equation}{#2#1#3}

\crefname{section}{Sec.}{Sec.}
\crefname{appendix}{Appendix}{Appendices}
\crefname{figure}{Fig.}{Figs.}

\crefname{definition}{Def.}{Defs.}
\crefname{prop}{Prop.}{Props.}
\crefname{lemma}{Lemma}{Lemmas}
\crefname{corollary}{Cor.}{Cors.}
\crefname{thm}{Theorem}{Theorems}
\crefname{remark}{Remark}{Remarks}

\crefname{ass}{Assumptions}{Assumptions}
\crefname{property}{Properties}{Properties}

\newcommand{\be}{\begin{equation}}
\newcommand{\ee}{\end{equation}}

\newcommand{\lb}{\left}
\newcommand{\rb}{\right}

\newcommand{\mc}{\mathcal}

\newcommand{\ms}{\mathscr}
\newcommand{\mf}{\mathfrak}
\newcommand{\bb}{\mathbb}

\newcommand{\eqsp}{\, ,\quad} 

\newcommand{\union}{\cup} 
\newcommand{\inter}{\cap} 

\newcommand{\abs}[1]{\lb\vert\, #1 \,\rb\vert}		

\newcommand{\cc}{\lambda} 

\newcommand{\Lie}{\pounds} 
\newcommand{\defn}{\mathrel{\mathop:}=} 

\renewcommand{\bar}{\overline}

\begin{document}

\setstretch{1.2}


\title{Stability of stationary-axisymmetric black holes in vacuum general relativity to axisymmetric electromagnetic perturbations}

\author{Kartik Prabhu}\email{kartikprabhu@cornell.edu}
\affiliation{Cornell Laboratory for Accelerator-based Sciences and Education (CLASSE)\\ Cornell University, Ithaca, NY 14853, USA}
\author{Robert M. Wald}\email{rmwa@uchicago.edu}
\affiliation{Enrico Fermi Institute and Department of Physics, The University of Chicago, Chicago, IL 60637, USA}

\begin{abstract}
We consider arbitrary stationary and axisymmetric black holes in general relativity in $(d +1)$ dimensions (with $d \geq 3$) that satisfy the vacuum Einstein equation and have a non-degenerate horizon. We prove that the canonical energy of axisymmetric electromagnetic perturbations is positive definite. This establishes that all vacuum black holes are stable to axisymmetric electromagnetic perturbations. Our results also hold for asymptotically deSitter black holes that satisfy the vacuum Einstein equation with a positive cosmological constant. Our results also apply to extremal black holes provided that the initial perturbation vanishes in a neighborhood of the horizon.
\end{abstract}

\maketitle
\tableofcontents


\section{Introduction}\label{sec:intro}

It is of considerable interest to determine the stability of stationary black hole solutions to Einstein equation. For a solution to be physically relevant, it is essential that sufficiently small perturbations not drive one away from that solution. The full nonlinear stability problem has been settled only for Minkowski spacetime \cite{Ch-Kl}. As a first step for other cases, it is important to analyze the stability of solutions to linearized perturbations. To establish the linear stability of a solution, one must show that all initial data for the linearized equations that are suitably regular and satisfy appropriate asymptotic conditions give rise to solutions that remain uniformly bounded and, further, decay at asymptotically late times to a stationary solution. On the other hand, a considerably weaker notion of linear stability that is much easier to analyze is {\em mode stability}, i.e., the nonexistence of suitably regular solutions that grow exponentially in time. 

For the case of gravitational perturbations satisfying the linearized Einstein equation, mode stability of Schwarzschild spacetime follows immediately from the form of the decoupled equations for the perturbations \cite{RW, Zerilli, IK-stab}. Recently, a complete proof of linear stability, including decay, of gravitational perturbations of \(4\)-dimensional Schwarzschild has been given by Dafermos, Holzegel and Rodnianski \cite{DHR}. However, their methods do not admit a straightforward extension to the Kerr case and, moreover, are special to the case of \(4\)-dimensions. Much less is known about the stability of general black holes in $d > 4$ dimensions, where there is a large variety of black hole solutions, some of which are known and/or believed to be unstable \cite{ER}.

As simpler problems than considering gravitational perturbations---which, nevertheless, should display many of its features---one could analyze the stability of vacuum black hole spacetimes to scalar perturbations satisfying the massless Klein-Gordon equation or to electromagnetic perturbations satisfying Maxwell's equations. In the case of a Kerr black hole in $4$-dimensions, complete results on stability and decay of scalar fields have been obtained \cite{FKSY, And, Tataru, DRS-stab}. For electromagnetic perturbations, boundedness and decay on a Schwarzschild black hole background has been shown by \cite{Blue} using the Maxwell energy-momentum tensor and by \cite{ABB} using a higher-derivative ``superenergy tensor'' (see also \cite{Pasq}). For a slowly rotating Kerr black hole, uniform energy bounds were established in \cite{AB}. However, these methods cannot be straightforwardly generalized to treat black holes in higher dimensions.

As a very significant simplification, one could consider the stability of black holes to axisymmetric perturbations. Here, by ``axisymmetric perturbations'' we mean the following. At infinity, the horizon Killing field $\chi^\mu$ will take the form of a linear combination of the stationary Killing field $t^\mu$ and rotational Killing fields \(\phi^\mu_\Lambda\) (with \(\Lambda \in \{1,2\ldots, k \}\))
\be
\chi^\mu = t^\mu + \sum_{\Lambda =1}^k \Omega^\Lambda \phi^\mu_\Lambda \, .
\label{axi}
\ee
We will further assume that the metric possesses a $t$-$\phi$ reflection isometry---as was proven to hold for vacuum solutions in \cite{SW-tphi}---so that the spacetime metric may be put in the form
\be
ds^2 = - N^2 dt^2 + 2 N_\Lambda dt d\phi^\Lambda + \Phi_{\Lambda \Theta} d \phi ^\Lambda d \phi^\Theta + \gamma_{ij} dx^i dx^j
\label{coord}
\ee
where, in these coordinates, we have $t^\mu = (\partial/\partial t)^\mu$ and $\phi^\mu_\Lambda = (\partial/\partial \phi^\Lambda)^\mu$.
By an ``axisymmetric perturbation,'' we mean a perturbation that is invariant under the action of \emph{all} \(\phi^\mu_\Lambda\) appearing in \cref{axi}. We do {\em not} require invariance under any axial Killing fields that may be present in the spacetime but are not associated with the rotation of the horizon. In particular, for a static black hole (where $\chi^a = t^a$), we do not place any symmetry restriction on the perturbation.

For axisymmetric gravitational perturbations, the canonical energy method of Hollands and Wald \cite{HW-stab} provides a general approach to analyzing stability. Positivity of canonical energy immediately implies mode stability \cite{HW-stab}, whereas failure of positivity implies that there exist perturbations that grow exponentially in time \cite{PW}. However, the expression for the canonical energy of gravitational perturbations is quite unwieldy, particularly since the linearized constraint equations must be imposed upon the variables appearing in the expression. It has not even been shown directly from the formula for canonical energy that the canonical energy is positive for perturbations of a $4$-dimensional Schwarzschild black hole, where it must be positive due to the known stability of these solutions (see above). On the other hand, certain black rings in $5$-dimensions can be shown to be unstable to axisymmetric perturbations \cite{FMR} by these methods \cite{SW-turning-pt}. However, in general, the stability of black holes to axisymmetric gravitational perturbations remains an open problem (see \cite{ER}). 

By contrast, it is straightforward to show the stability of all black holes to axisymmetric massless Klein-Gordon perturbations.
For a scalar field, $\varphi$, the energy-momentum tensor is given by 
\be
T_{\mu \nu} (\varphi) = \nabla_\mu \varphi  \nabla_\nu \varphi - \frac{1}{2} g_{\mu \nu} \nabla^\alpha \varphi \nabla_\alpha \varphi 
\label{tpcoord}
\ee
The energy of a scalar field associated with this energy-momentum tensor is 
\be
{\ms E}_\varphi \defn \int_\Sigma T_{\mu \nu} (\varphi) t^\mu u^\nu
\label{pse}
\ee
where we may take $\Sigma$ to be a hypersurface of constant $t$ in the coordinates \cref{coord}, and $u^\mu$ denotes the unit normal to $\Sigma$. It should be noted that---unlike the case of a Maxwell field (see below)---for a Klein-Gordon field, the energy \cref{pse} defined via the energy-momentum tensor agrees with the canonical energy. For axisymmetric perturbations, we have $\phi^\mu_\Lambda \nabla_\mu \varphi = 0$ and we may therefore replace 
$t^\mu$ by $N u^\mu$ in \cref{pse}. We thereby obtain the manifestly positive definite expression
\be
{\ms E}_\varphi = \frac{1}{2} \int_\Sigma N \left[(u^\mu \nabla_\mu \varphi)^2 + D^a \varphi D_a \varphi \right] \, .
\ee
where $D_a$ is the derivative operator on $\Sigma$.
Conservation of $\ms E_\varphi$ then immediately implies mode stability. Indeed, since $\ms E_\varphi$ has the character of a Sobolev norm and higher Sobolev-like norms can be obtained from the energy of time derivatives of $\varphi$, it should be straightforward\footnote{Difficulties caused by the vanishing of $N$ on the horizon can be dealt with (in the case of a non-degenerate horizon) by the ``trick'' used in \cite{KW-stab} or the methods described in  \cite{Daf-Rod-lec}.} to show boundedness of axisymmetric perturbations $\varphi$ off of an arbitrary black hole background. However, we shall not consider any boundedness and decay results here. Note that the positivity of $\ms E_\varphi$ holds for axisymmetric $\varphi$ in any spacetime of the form \cref{coord}, i.e., it is not necessary to assume that the background black hole satisfies Einstein equation.

The case of axisymmetric electromagnetic perturbations is intermediate between the gravitational and scalar cases. The electromagnetic energy-momentum tensor of the Maxwell field $F_{\mu \nu} = 2 \nabla_{[\mu} A_{\nu]}$ is given by
\be\label{eq:T-EM}
	T_{\mu\nu}(F) = \frac{1}{4\pi} \lb[F_{\mu\lambda} F_\nu{}^\lambda - \tfrac{1}{4} g_{\mu\nu} F_{\rho\lambda}F^{\rho\lambda}\rb] 
\ee
The electromagnetic energy associated with this energy-momentum tensor is
\be\label{eq:E-T1}
{\ms E}_F = \int_\Sigma T_{\mu \nu} (F) t^\mu u^\nu
\ee
where we have inserted a subscript ${F}$ to distinguish this energy from the canonical energy \(\ms E\), defined in \cref{sec:em-pert} below.
The electric field on $\Sigma$ is defined by $E_\mu = u^\nu F_{\nu \mu}$. In terms of $E_a$ and the pullback, $A_a$, of $A_\mu$ to $\Sigma$, we have 
\be\label{eq:E-T}
	\ms E_F =\frac{1}{8\pi} \int_\Sigma N E_a E^a + 2N (D_{[a}A_{b]})(D^{[a}A^{b]}) + 4 N^a E^b D_{[a} A_{b]}
\ee
where $N^a$ is the shift vector of $t^\mu$. In an ergoregion, we have $N^a N_a > N^2$, so the right-hand-side of \cref{eq:E-T} need not be pointwise positive even for axisymmetric electromagnetic fields. Since $E_a$ must satisfy the constraint $D^a E_a = 0$, it is not immediately obvious by inspection whether the integral on the right-hand-side of \cref{eq:E-T} can be made negative for axisymmetric electromagnetic fields. It was shown in Appendix B.2 of \cite{PW} that one can find spacetimes of the form \cref{coord} for which the total electromagnetic energy can be made negative. However, it is not obvious whether this is possible for black holes that are solutions to the vacuum Einstein equation. As a by-product of our analysis, we will show in \cref{sec:non-positivity} that, in fact, $\ms E_F$ can be made negative for axisymmetric electromagnetic perturbations of any rotating black hole background that satisfies the vacuum Einstein equation, such as Kerr.

The main purpose of this paper is to prove that, nevertheless, the {\em canonical energy}, $\ms E$, of axisymmetric electromagnetic perturbations is positive definite\footnote{The canonical energy can be degenerate on stationary solutions (see \cref{prop:E-props} below), so we must factor out these stationary, degenerate solutions (if any exist) to obtain positive definiteness.} on all stationary black hole backgrounds with a non-degenerate horizon that satisfy the vacuum field equations. The canonical energy, $\ms E$, differs from $\ms E_F$ by a boundary term\footnote{This boundary term was erroneously omitted in Appendix B.2 of \cite{PW}.} on the bifurcation surface of the horizon, and this boundary term is essential to obtain positivity. The positive definiteness and conservation of $\ms E$ ensure mode stability. Since $\ms E$ provides a Sobolev-like norm and higher Sobolev-like norms can be obtained from the canonical energy of time derivatives of solutions, we believe that boundedness of the electromagnetic field tensor could also be proven, but we shall not attempt to investigate this issue here.

In \cref{sec:bg}, we review the properties of the stationary-axisymmetric background black hole spacetimes that we will consider. In \cref{sec:em-pert} we define the canonical energy for electromagnetic perturbations and establish some of its key properties. In \cref{sec:axiEM}, we restrict to axisymmetric perturbations. We split the canonical energy into ``kinetic'' and ``potential'' parts according to behavior under the $t$-$\phi$ reflection isometry of the background spacetime. We show that the kinetic energy is positive-definite, even if the black hole background does not satisfy Einstein equation. We then prove our main result showing that the potential energy for axisymmetric electromagnetic perturbations is positive whenever the vacuum Einstein equation holds. 

In \cref{sec:U-pos-other}, we show that for perturbations with vanishing charge the canonical energy can be put into other positive definite forms in $3+1$ and $4+1$ spacetime dimensions. In particular, in \(4\)-spacetime dimensions, the potentials for the electric and magnetic fields can be combined into a complex scalar potential \(\Psi\) for the self-dual part of the electromagnetic field strength. We show that \(\Psi\) satisfies a ``wave-like" equation in spacetime (\cref{eq:Psi-eom}) and that the canonical energy can be written in terms of an ``effective energy-momentum tensor'' for \(\Psi\) (\cref{eq:T-eff}). Finally, in \cref{sec:non-positivity}, we show that $\ms E_F$ can always be made negative for axisymmetric perturbations of a rotating black 
hole satisfying the vacuum Einstein equation.\\

In this paper, lower case Greek indices will be used to denote tensors on spacetime, e.g., $g_{\mu \nu}$ denotes the spacetime metric and $t^\mu$ denotes the timelike Killing field of the background black hole. Lower case Latin indices will be used to denote tensors on the initial data surface $\Sigma$, e.g., $h_{ab}$ denotes the metric on $\Sigma$. The spacetime derivative operator of the background black hole will be denoted as $\nabla_\mu$; the background derivative operator on $\Sigma$ will be denoted as $D_a$. Capital Greek indices will label the axial Killing fields, $\phi^a_\Lambda$. As usual, spacetime and space indices will be raised and lowered using the appropriate metrics, $g_{\mu \nu}$ and $h_{ab}$. Capital Greek indices will be raised and lowered using $\Phi_{\Lambda \Theta}$, defined in \cref{Phi} below. The rest of our conventions follow those of Wald \cite{Wald-book}.

\section{Background stationary-axisymmetric black hole spacetimes}\label{sec:bg}

\begin{figure}[h!]
	\centering
	\includegraphics[width=0.5\textwidth]{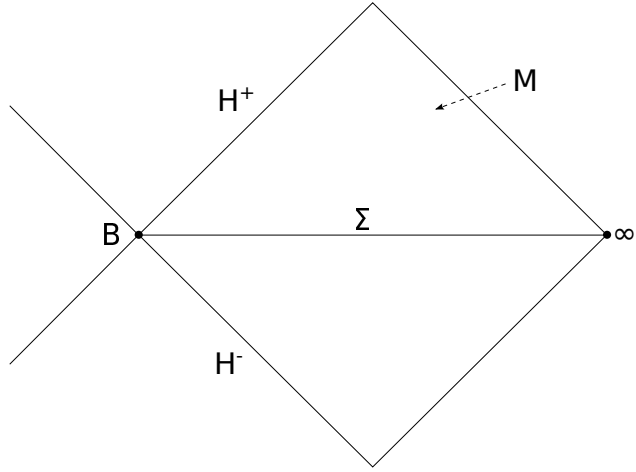}
	\caption{Carter-Penrose diagram of the black hole spacetime \((M,g)\).}\label{fig:spacetime}
\end{figure}

We consider an asymptotically flat \((d+1)\)-dimensional spacetime with $d \geq 3$, containing a static or stationary-axisymmetric black hole spacetime \(( M, g)\) shown in \cref{fig:spacetime}, with a bifurcate Killing horizon \(  H \defn  H^+ \union  H^- \), and bifurcation surface \(  B \defn  H^+ \inter  H^- \). We assume that \(B\) is compact, but we do not assume any further restrictions on its topology\footnote{In particular, we need not assume that $B$ is connected, although for notational convenience (so that we do not have multiple horizon Killing fields), we will assume that this is the case.}. Let $t^\mu$ denote the time translation Killing field and let $\phi^\mu_\Lambda$ denote the axial Killing fields associated with the horizon rotation (see \cref{axi}).
We assume that the spacetime possesses a $t$-$\phi$ reflection isometry (as has been proven to hold for vacuum solutions \cite{SW-tphi}), so that the spacetime metric may be put in the form \cref{coord}.
Let $\Sigma$ denote a $t$-$\phi$ reflection invariant surface (i.e., a $t = {\rm const.}$ surface in the coordinates of \cref{coord}). We assume that $\Sigma$ has one asymptotically flat end (with asymptotic conditions given by \cref{eq:falloff-bg} below) and that inside a large sphere $S_\infty$ in the asymptotic region, $\Sigma$ is a compact manifold with boundary $S_\infty \union B$. Let $u^\mu$ denote the future-directed unit normal to \(\Sigma\). We decompose $t^\mu$ into its normal and tangential parts relative to $\Sigma$, referred to as the \emph{lapse}, \(N = - u_\mu t^\mu\), and \emph{shift}, \(N^a\), on \(\Sigma\).

Let \( h_{ab} \) denote the induced metric on $\Sigma$, let \(K_{ab}\) denote the extrinsic curvature of $\Sigma$, and define
\be
\pi^{ab} \defn K^{ab}-K~h^{ab} \, .
\ee
On the bifurcation surface \( B\), we introduce a unit normal vector \(r^a\) (pointing into \(\Sigma\)). Note that $N\vert_B =0$. The asymptotic flatness conditions on our stationary black hole are that there exist coordinates \( (x_1, \ldots, x_d) \) on \(\Sigma\) such that
\be\label{eq:falloff-bg}\begin{split}
	h_{ab} \sim \delta_{ab} + O(1/\rho^{d-2}) \eqsp \pi^{ab} \sim O(1/\rho^{d-1}) \\
	N \sim 1 + O(1/\rho^{d-2}) \eqsp N^a \sim O(1/\rho^{d-1})
\end{split}\ee\
where \( \rho \defn \lb( x_1^2 + \ldots + x_d^2 \rb)^{1/2} \) near infinity. In addition, all \(n\)\textsuperscript{th} derivatives of the above quantities (with respect to asymptotically Cartesian coordinates) are required to fall off faster by an additional factor of \(1/\rho^n\). The asymptotic conditions on the lapse and shift ensure that $t^\mu$ goes to an asymptotic time-translation at infinity.

It is convenient to view the labeling index, $\Lambda$, of the axial Killing fields $\phi^a_\Lambda$ as an abstract index associated with a vector space \(\bb V\) of axial Killing fields (see \cite{PW}). At each $x \in \Sigma$ where the axial Killing fields are linearly independent, we can then define a positive definite inverse metric $\Phi_{\Lambda\Theta}(x)$ on $\bb V$ by
\be
\Phi_{\Lambda\Theta}(x) \defn h_{ab}(x) \phi^a_\Lambda\phi^b_\Theta  
\label{Phi}
\ee
We use $\Phi_{\Lambda\Theta}$ and its inverse, $\Phi^{\Lambda\Theta}$ to lower and raise the indices \(\Lambda, \Theta \ldots\) in the vector space of the axial Killing fields. It is important to note that \(D_a\Phi_{\Lambda\Theta} \neq 0\), so, while \(\phi^a_\Lambda\) satisfies Killing's equation, \(\phi^{a\Lambda}\) does not. The $(d-k)$-dimensional surface orthogonality of \(\phi^a_\Lambda\) within $\Sigma$ together with Killing's equation implies that 
\be\label{eq:Dphi1}
	D_{a} \phi_{b\Lambda}  = - \Phi^{\Theta \Xi} \phi_{\Theta[a} D_{b]}\Phi_{\Lambda \Xi} 
\ee
This implies that
\be
 D_{[a} \phi_{b]}^\Lambda   = 0
 \label{eq:Dphi}
\ee

It will be convenient for later use to introduce the quantity
\be\label{eq:det-Phi}
\Phi \defn {\rm det} (\Phi_{\Lambda\Theta})
\ee
where the determinant is taken with respect to the $k$-form $\eta_{\Lambda_1 \dots \Lambda_k}$ on \(\bb V\) with $\eta_{12 \dots k}  = 1$ in the original basis of axial Killing fields (normalized so that a rotational period is $2 \pi$). It also is convenient to introduce the quantity
\be\label{eq:X-choice}
	X_a{}^\Theta{}_\Lambda = -\tfrac{1}{2} \Phi^{\Theta \Xi} D_a \Phi_{\Lambda \Xi}
\ee
so that, in view of \cref{eq:Dphi1}, we have
\be
D_{a} \phi_{b\Lambda}  = 2 \phi_{\Theta [a} {{X_{b]}}^\Theta}_\Lambda 
\ee

In view of the ($t$-$\phi$)-reflection isometry about $\Sigma$, the shift \(N^a\) and momentum \(\pi^{ab}\) are of the form (see \cite{PW})
\begin{subequations}\begin{align}
	N^a &= N^\Lambda\phi^a_\Lambda \\
	\pi^{ab} &= 2 \pi^{\Lambda(a}\phi^{b)}_\Lambda 
\end{align}\end{subequations}
with $\pi^\Lambda_a\phi^a_\Theta = 0$. Here $N^\Lambda = \Phi^{\Lambda \Theta} N_\Theta$, where $N_\Theta$ is the quantity appearing in \cref{coord}. 

The vacuum Einstein equation requires that the initial data \((\pi^{ab}, h_{ab})\) satisfy the constraint equations. For a stationary, axisymmetric spacetime with a $t$-$\phi$ reflection isometry, the Hamiltonian and momentum constraints become (see \cite{PW}):
\begin{subequations}\begin{align}
	R & = 2\pi_a^\Lambda \pi^a_\Lambda \label{eq:H-constraint}\\
	D_a\pi^a_\Lambda & = 0 \label{eq:diff-constraint}
\end{align}\end{subequations}
where $R$ is the scalar curvature of the metric $h_{ab}$ on $\Sigma$ and $D_a$ is the derivative operator on $\Sigma$. 
The ADM time evolution equations (see, e.g., Sec.E.2 of \cite{Wald-book} for \(d=3\) and Sec.VI.6 of \cite{CB-book} for general \(d\)) for stationary solutions also impose restrictions on the initial data.
For spacetimes of the type we consider, the ADM evolution equations yield
\begin{subequations}\label{eq:evol-bg-tphi}\begin{align}
NR_{ab} & = D_aD_b N -2 N\lb( \pi_a^\Lambda\pi_{b\Lambda} - \pi_c^\Lambda \pi^{c\Theta} \phi_{a\Lambda}\phi_{b\Theta} \rb) \label{eq:Ric-id} \\
	D_a N^\Lambda & = - 2 N\pi_a^\Lambda \label{eq:Dshift}
\end{align}\end{subequations}

Finally, we note that the relation $D^b D_b \phi^a_\Lambda = - {R^a}_b \phi^b_\Lambda$ (which holds for any Killing field) may be combined with \cref{eq:Dphi1} to yield
\be
	R_{ab}\phi^b_\Lambda =  \phi_{a \Theta} D^b X_b{}^{\Theta}{}_{\Lambda}
\ee
where $X_b{}^{\Theta}{}_{\Lambda}$ was defined in \cref{eq:X-choice} above.
When combined with \cref{eq:Ric-id}, we obtain the following very useful relation
\be\label{eq:useful}
	D^a (N X_a{}^{\Theta}{}_{\Lambda}) = 2 N \pi_a^\Theta \pi^a_\Lambda \, .
\ee
Equivalently, we have
\be
\label{eq:DX-id}
	D^a (N X_a{}^{\Lambda \Theta})  = 2 N X_a{}^{\Lambda\Xi} X^{a \Theta}{}_\Xi + 2 N \pi_a^\Lambda \pi^a{}^\Theta  \, .
\ee

\section{Electromagnetic perturbations and canonical energy}\label{sec:em-pert}

We wish to consider electromagnetic perturbations of the black hole spacetimes described in the previous section. The electromagnetic field is represented by a connection in a principal $U(1)$ bundle over spacetime, but linearized perturbations must be described by a trivial bundle, so we may represent the electromagnetic field by a smooth vector potential \(A_\mu\) on spacetime, which is unique up to smooth gauge transformations $A_\mu \mapsto A_\mu - \nabla_\mu \xi$. We require that $A_\mu$ go to zero at spatial infinity as
\be\label{eq:falloff-EM}
 A_\mu \sim O(1/\rho^{d-2})
\ee
where \(\rho\) is the asymptotic radial coordinate introduced below \cref{eq:falloff-bg}. All \(n\)\textsuperscript{th} spacetime derivatives of $A_\mu$ are required to fall off faster by an additional factor of \(1/\rho^n\). In this section, we will not assume that the electromagnetic field is axisymmetric.

The Maxwell Lagrangian is
\begin{equation}
	L_{\mu_1 \dots \mu_{d+1}}
		= - \frac{1}{16\pi} F^{\nu\lambda} F_{\nu\lambda} ~\varepsilon_{\mu_1 \dots \mu_{d+1}}
\label{maxlag}
\end{equation}
where  $F_{\mu\nu} \defn 2 \nabla_{[\mu}A_{\nu]}$ and $\varepsilon_{\mu_1 \dots \mu_{d+1}}$ is the volume element associated with the spacetime metric. The equations of motion obtained from \cref{maxlag} are Maxwell's equations
\be\label{eq:maxwell}
	\nabla^\mu F_{\mu\nu} = 0
\ee
The symplectic potential obtained from \cref{maxlag} is
\be
\theta_{\mu_1 \dots \mu_d} =  - \frac{1}{4\pi} F^{\nu\lambda} \delta A_\lambda ~ \varepsilon_{\nu \mu_1 \dots \mu_d}
\ee
and the symplectic current is
\be\label{symcur}
\omega_{\mu_1 \dots \mu_d} = - \frac{1}{4\pi} \left[(\delta_1 F^{\nu\lambda}) (\delta_2 A_\lambda) - (\delta_2 F^{\nu\lambda}) (\delta_1 A_\lambda) \right] \varepsilon_{\nu \mu_1 \dots \mu_d}
\ee
Here $A_\mu$ denotes the background vector potential and $\delta A_\mu$ denotes the perturbed vector potential. We have $d {\bf \omega} = 0$ whenever $\delta_1 A_\mu$ and $\delta_2 A_\mu$ satisfy Maxwell's equations. Since the background electromagnetic field vanishes\footnote{Since Maxwell's equations are linear, the background electromagnetic field does not appear in the expressions for the symplectic current and canonical energy in any case.} and we are only interested in perturbed quantities, we will drop the $\delta$ in all further expressions, e.g., in all formulas below, $A_\mu$ will denote the perturbed vector potential.

The initial data for the electromagnetic field on \(\Sigma\) is given by \(\alpha \defn (E_a,  A_a)\), where $E_a$ is the \emph{electric field} defined by 
\be
E_\mu = u^\nu F_{\nu\mu}
\ee
and $A_a$ is the pullback of \(A_\mu\) to $\Sigma$. Maxwell's equations \cref{eq:maxwell} are equivalent to the Gau\ss\ constraint
\be\label{eq:gauss}
	D^a E_a = 0 
\ee
together with evolution equations for the initial data\footnote{The evolution equations in Appendix B.2 \cite{PW} have the incorrect sign for the shift vector which, however, does not affect the results.}
\be\label{eq:evol-EM}\begin{split}
	\dot A_a & = D_a(t^\mu A_\mu) + NE_a - 2N^bD_{[a}A_{b]} \\
	\dot E_a & = 2D^b\lb( ND_{[b}A_{a]} + N_{[b}E_{a]}\rb) 
\end{split}\ee
The Maxwell energy-momentum tensor is given by \cref{eq:T-EM} and the corresponding energy, $\ms E_F$, is given by \cref{eq:E-T1}. As already noted in the Introduction, $\ms E_F$ can be written in terms of \( \alpha = (E_a,  A_a)\) as \cref{eq:E-T}.

The symplectic form $\Omega$ on the space of initial data is obtained by integrating \cref{symcur} over $\Sigma$
\be\label{eq:symp-defn}
\Omega (\alpha_1, \alpha_2) \defn \int_\Sigma {\bf \omega} = \tfrac{1}{4\pi}\int_\Sigma (E_1^a A_{2a} - A_{1a} E_2^a) \, ,
\ee
where the natural volume element on $\Sigma$ is understood. This integral converges by virtue of our asymptotic conditions \cref{eq:falloff-EM} on $A_\mu$ except in $(3+1)$-dimensions, where there could be a logarithmic divergence in $\rho$. This difficulty can be eliminated by imposing slightly stronger fall-off conditions on the odd parity part of $E_a$ and the even parity part of $A_a$ (analogous to the Regge-Teitelboim conditions on the asymptotic behavior of the metric at spatial infinity \cite{RT}). However, we shall not impose any such additional conditions here because the canonical energy integral (see \cref{eq:E-symp-defn} below) will converge without the imposition of additional conditions on account of the extra derivative appearing in that formula.
Since $d {\bf \omega} = 0$, the symplectic form is conserved in the sense that the integral of \cref{symcur} takes the same value on any Cauchy surface for the exterior of the black hole.

The \emph{canonical energy} is obtained from $\Omega$ by replacing the solution $A_{2\mu}$ by the solution $\Lie_t A_{2 \mu}$. Viewed as a bilinear form on initial data \(\alpha = (E_a,  A_a)\), we have
\be\label{eq:E-symp-defn}
	\ms E(\alpha_1, \alpha_2) = \frac{1}{4\pi}\int_\Sigma (E_1^a \dot A_{2a} - A_{1a} \dot E_2^a) 
\ee
where $\dot A_{2a}$ and $\dot E_{2a}$ are given by \cref{eq:evol-EM}. The key properties of $\ms E$ are given by the following proposition:

\begin{prop}\label{prop:E-props} The canonical energy $\ms E(\alpha_1, \alpha_2)$ satisfies the following properties: (i) $\ms E$ is conserved. (ii) $\ms E$ is symmetric
\be
{\ms E} (\alpha_1, \alpha_2) = {\ms E} (\alpha_2, \alpha_1) \, .
\ee
(iii) As a quadratic form, \(\ms E\) can be degenerate only on physically stationary solutions, i.e., solutions for which $\Lie_t F_{\mu \nu} = 0$.
(iv) For axisymmetric perturbations, $\ms E$ is invariant under Maxwell gauge transformations.
\begin{proof}
Conservation of $\ms E$ (property (i))  follows immediately from conservation of $\Omega$. 

The symmetry of $\ms E$ (property (ii))
can be seen as follows \cite{HW-stab}: In differential forms notation, we have
\be
\Lie_t {\bf \omega} = d(t \cdot {\bf \omega}) + t \cdot d {\bf \omega} = d(t \cdot {\bf \omega}) 
\ee
since $d {\bf \omega} = 0$. Thus, when integrated over $\Sigma$, the right side yields only boundary terms. These boundary terms vanish due to our asymptotic conditions \cref{eq:falloff-EM} and the fact that $t^\mu$ is tangent to $B$. Symmetry of $\ms E$ then follows immediately.

The non-degeneracy of $\ms E$ (property (iii)) can be seen as follows. Suppose that $\alpha = (E_a, A_a)$ is a degeneracy of $\ms E$, i.e., suppose that
\be
\ms E (\alpha', \alpha) = \tfrac{1}{4\pi}\int_\Sigma (E'^a \dot A_a - A'_a \dot E^a)  = 0 
\ee
for all $\alpha'$. Since $A'_a$ can be chosen to be an arbitrary smooth vector field of compact support, we immediately obtain $\dot E^a = 0$. On the other hand, we can choose 
\be
E'^a = D_b U'^{ab}
\ee
where $U'^{ab} = U'^{[ab]}$ is an arbitrary antisymmetric tensor field of compact support. We thereby obtain $D_{[a} \dot A_{b]} = 0$. These conditions are equivalent to $\Lie_t F_{\mu \nu} = 0$.

To prove gauge invariance (property (iv)), we must show that $\ms E (\alpha_1, \alpha_2) = 0$ whenever $\alpha_2$ is of the form $\alpha_2 = (0, -D_a \xi)$. Substituting this in \cref{eq:E-symp-defn}, we obtain
\be
	\ms E(\alpha_1, \alpha_2) = - \tfrac{1}{4\pi}\int_\Sigma E_1^a D_a {\dot \xi} \, .
\ee
Since $D_a E_1^a = 0$, we can convert the integral over $\Sigma$ into a boundary integral at infinity and at $B$. The boundary integral at infinity vanishes by our asymptotic conditions, and---taking account of \cref{axi} and the fact that $\chi^\mu$ vanishes at $B$---the boundary integral at $B$ vanishes by axisymmetry.
\end{proof}
\end{prop}

\begin{remark} Although every degeneracy of $\ms E$ corresponds to a physically stationary solution, stationary solutions with nonvanishing charge will not, in general, be degeneracies of $\ms E$.
\end{remark}
\medskip

\begin{remark} The restriction to axisymmetry is not necessary to prove gauge invariance of the canonical energy $\ms E'$ defined with respect to the horizon Killing field $\chi^\mu$ rather than $t^\mu$. In the proof of property (iv) above, the boundary term at $B$ will vanish without any assumption of axisymmetry. However, in $3+1$ dimensions, one would have to impose the slightly stronger asymptotic fall-off conditions mentioned below \cref{eq:symp-defn} in order to be guaranteed of getting a vanishing contribution from the boundary term at infinity.
\end{remark}
\medskip

Since the only degeneracies of $\ms E$ are physically stationary solutions and since stationary solutions are manifestly stable, mode stability will be proven if $\ms E (\alpha, \alpha) \geq 0$ for all $\alpha$. 
Substituting from \cref{eq:evol-EM} into \cref{eq:E-symp-defn} and using \cref{eq:gauss}, we obtain (after various integrations-by-parts, using our fall-off conditions \cref{eq:falloff-EM} at infinity) the following explicit expression for the canonical energy:
\be\label{eq:E-real}
\ms E \defn \ms E(\alpha, \alpha) = \tfrac{1}{4\pi}\int_\Sigma N E_a^2 + 2N (D_{[a}A_{b]})^2 + 4 N^a E^b D_{[a} A_{b]} - \tfrac{1}{4\pi} \int_B 2 r^a E_a N^b A_b
\ee
Thus, the canonical energy \(\ms E\) \cref{eq:E-real} differs\footnote{The factor of $1/2$ appearing on the left side of \cref{eet} occurs because, following \cite{HW-stab}, we have normalized $\ms E$ so that for a one-parameter family of solutions $A_\mu(\lambda)$, it is the second derivative of energy with respect to $\lambda$. The factor of $1/2$ is the Taylor coefficient of this term.} from the energy \(\ms E_F\) obtained from the Maxwell energy-momentum tensor \cref{eq:E-T} by a boundary term on the bifurcation surface\footnote{As previously mentioned, this boundary term was erroneously omitted in Appendix B.2 \cite{PW}.}
\be\label{eet}
\tfrac{1}{2} \ms E =  \ms E_F - \tfrac{1}{4\pi} \int_B r^a E_a N^b A_b \, .
\ee
This boundary term vanishes for a static black hole (since \(N^a = 0\)), and for perturbations supported away from \(B\) on a stationary-axisymmetric black hole. However, as we shall see, it will be important to keep this term to show positivity of the canonical energy for perturbations on a stationary-axisymmetric background which do not vanish on \(B\).

When the background black hole is static ($N^a = 0$), we get a manifestly positive-definite energy
\be
	\tfrac{1}{2} \ms E = \ms E_F = \tfrac{1}{8\pi}\int_\Sigma N E_a^2 + 2N (D_{[a}A_{b]})^2
\ee
Thus, any static vacuum black hole background (\emph{not} necessarily satisfying Einstein equation) is stable to (\emph{not} necessarily axisymmetric) electromagnetic perturbations \cite{HW-stab, PW}.

However, the situation is quite different for the case of a rotating black hole. For a rotating black hole, we still have $N=0$ on $B$ but we now have $N^a \neq 0$ on $B$, so there is always an {\it ergoregion} outside the horizon where $N^a N_a > N^2$. It is easy to find non-axisymmetric perturbations of compact support in the ergoregion that make $\ms E$ and $\ms E_F$ negative. For axisymmetric pertubations, if the spacetime is not required to satisfy Einstein equation, then it also is easy to construct examples (by choosing $N^a$ sufficiently large) for which $\ms E$ and $\ms E_F$ can be made negative (see Appendix B.2 \cite{PW}). Note that axisymmetric perturbations with $\ms E < 0$ will grow exponentially with time \cite{PW}, so such black holes are unstable to electromagnetic perturbations. However, if the spacetime satisfies Einstein equation, it is far from clear whether such examples are possible. In the next section, we will prove that $\ms E$ is positive whenever the background spacetime satisfies the vacuum Einstein equation. Thus, all black holes that satisfy the vacuum Einstein equation are stable to axisymmetric electromagnetic perturbations. Nevertheless, in \cref{sec:non-positivity}, we will show that for black holes that satisfy the vacuum Einstein equation, although $\ms E \geq 0$ for all axisymmetric electromagnetic perturbations, we can always find axisymmetric electromagnetic perturbations that make $\ms E_F < 0$.

\section{Positivity of $\ms E$ for axisymmetric perturbations}\label{sec:axiEM}

Consider initial data $(E_a, A_a)$ on $\Sigma$ that is axisymmetric, i.e., $\Lie_{\phi_\Lambda} E_a = \Lie_{\phi_\Lambda} A_a = 0$ for all $\phi^a_\Lambda$ appearing in \cref{axi}.
We decompose $E_a$ and $A_a$ into their $\phi$-reflection odd and even parts as
\begin{subequations}\begin{align}
	E_a & = E^\Lambda \phi_{\Lambda a} + \tilde E_a \\
	A_a & = A^\Lambda \phi_{\Lambda a} + \tilde A_a
\end{align}\end{subequations}
with $ \tilde E^a  \phi_{\Lambda a} = \tilde A^a  \phi_{\Lambda a} = 0$. Note that $A^\Lambda$ is gauge invariant with respect to gauge transformations that preserve $\Lie_{\phi_\Lambda} A_a = 0$. Note also that 
\be
A_a \phi^a_\Lambda = A^\Theta \phi_{\Theta a} \phi^a_\Lambda = A^\Theta \Phi_{\Theta \Lambda} = A_\Lambda \, .
\ee
The \(t\)-\(\phi\)-reflection even and odd parts of the initial data are, respectively,
\begin{subequations}\label{eq:PQ-decomp}\begin{align}
	P & = \begin{pmatrix} E^\Lambda \phi_{\Lambda a} , \tilde A_a \end{pmatrix} \\
	Q & = \begin{pmatrix}\tilde E_a, - A^\Lambda \phi_{\Lambda a} \end{pmatrix} 
\end{align}\end{subequations}
Note that \cref{eq:gauss} restricts only the \(t\)-\(\phi\)-reflection-odd data
\be\label{eq:polar-condns}
	D^a \tilde E_a = 0 
\ee

Under the decomposition \cref{eq:PQ-decomp} the canonical energy \(\ms E\) splits into independent parts which depend only on the \(t\)-\(\phi\)-reflection-even and -odd parts of the initial data. We call the reflection even part\footnote{Note that this reverses the terminology used for gravitational perturbations \cite{PW}; see Appendix B.2 \cite{PW}.} the \emph{kinetic energy} \(\ms K\) and the reflection odd part the \emph{potential energy} \(\ms U\). Thus, we have
\be
	\ms E= \ms K + \ms U 
\ee
with
\be
	\ms K \defn \ms E[(P, Q=0)] 
\ee
\be
	\ms U \defn \ms E[(P=0, Q)] \, .
\ee

The kinetic energy is given by the manifestly positive expression
\be\label{eq:KE}
	\ms K = \tfrac{1}{4\pi}\int_\Sigma N (E_\Lambda)^2 + 2 N (D_{[a}\tilde A_{b]})^2 
\ee
In the above expression the square of any tensor quantity is taken with respect to the appropriate metric, that is, \((E_\Lambda)^2 \defn \Phi^{\Lambda \Theta} E_\Lambda E_\Theta \) and \((D_{[a}\tilde A_{b]})^2 \defn h^{ac} h^{bd} D_{[a}\tilde A_{b]} D_{[c}\tilde A_{d]}\). We shall use this convention henceforth to avoid proliferation of indices. From \cref{eq:KE} we see that the kinetic energy is always positive for any black hole background, without the need to impose Einstein equation on the background.

The potential energy is given by
\be
	\ms U =  \tfrac{1}{4\pi}\int_\Sigma N\tilde E_a^2 + N \lb( D_{a}A_\Lambda \rb)^2  - 2 N^\Lambda \tilde E^b D_{b} A_\Lambda - \tfrac{1}{4\pi} \int_B 2 r^a \tilde E_a N^\Lambda A_\Lambda 
	\label{eq:U-decomp}
\ee
As remarked at the end of the previous section, there exist black hole spacetimes on which $\ms U$ fails to be positive for all $(E_a,  A_a)$ \cite{PW}. We now show that $\ms U \geq 0$ when the black hole background satisfies the vacuum Einstein equation.

To begin, we combine the last two terms in the expression \cref{eq:U-decomp} to obtain
\be\label{eq:IBP1}
- \int_\Sigma 2 N^\Lambda D_{a}(\tilde E^a A_\Lambda) - \int_B 2 r^a \tilde E_a N^\Lambda A_\Lambda = \int_\Sigma 2 (D_a N^\Lambda) \tilde E^a A_\Lambda 
\ee
Next, we substitute from \cref{eq:Dshift} to obtain
\be\begin{split}
\ms U &=  \tfrac{1}{4\pi}\int_\Sigma N \left[\tilde E_a^2 + ( D_{a}A_\Lambda)^2 - 4\pi^\Lambda_a \tilde E^a A_\Lambda \right]  \\
	&= \tfrac{1}{4\pi}\int_\Sigma N \left[\left(\tilde E_a - 2\pi^\Lambda_a  A_\Lambda \right)^2 + ( D_{a}A_\Lambda)^2 - 4(\pi^\Lambda_a  A_\Lambda)^2 \right]
\end{split}\ee
Using  \cref{eq:DX-id}, we get
\be\label{iden1}
-4 N(\pi^\Lambda_a  A_\Lambda)^2 = -4N \pi^{a\Lambda} \pi_a^\Theta A_\Lambda A_\Theta = -2 D^a (N X_a{}^{\Lambda \Theta})A_\Lambda A_\Theta + 4 N X_a{}^{\Lambda\Xi} X^{a \Theta}{}_\Xi A_\Lambda A_\Theta
\ee
where $X^{a \Theta}{}_\Xi$ was defined in \cref{eq:X-choice} above.
On the other hand, we have
\be\label{iden2}\begin{split}
 -2 D^a (N X_a{}^{\Lambda \Theta})A_\Lambda A_\Theta &= -2D^a (N X_a{}^{\Lambda \Theta} A_\Lambda A_\Theta) + 2 N X_a{}^{\Lambda \Theta} D^a (A_\Lambda A_\Theta) \\
	&= -2D^a (N X_a{}^{\Lambda \Theta} A_\Lambda A_\Theta) + 4 N X_a{}^{\Lambda \Theta} A_\Theta D^a A_\Lambda
\end{split}\ee
The integral over $\Sigma$ of first term on the right side of \cref{iden2} vanishes. We thereby obtain
\be\label{finpos}\begin{split}
\ms U & = \tfrac{1}{4\pi}\int_\Sigma N \left[\left(\tilde E_a - 2\pi^\Lambda_a  A_\Lambda \right)^2 + ( D_{a}A_\Lambda)^2 + 4 X_a{}^{\Lambda \Theta} A_\Theta D^a A_\Lambda + 4 X_a{}^{\Lambda\Xi} X^{a \Theta}{}_\Xi A_\Lambda A_\Theta \right] \\
	& = \tfrac{1}{4\pi}\int_\Sigma N \left[\left(\tilde E_a - 2\pi^\Lambda_a  A_\Lambda \right)^2 + \left( D_{a}A_\Lambda + 2 X_a{}^\Theta{}_\Lambda A_\Theta \right)^2 \right]
\end{split}\ee
Thus, $\ms U$ takes a manifestly positive form, as we desired to show.\\

\begin{remark}[Asymptotically deSitter backgrounds]\label{rem:dS}
Our results generalise to stationary-axisymmetric black holes in with a $t$-$\phi$ reflection isometry in asymptotically deSitter background spacetimes\footnote{Here, by ``asymptotically deSitter'' we mean a spacetime with a structure like that shown in \cref{fig:spacetime} above, but with ``$\infty$'' replaced by the bifurcation surface, $B'$, of another Killing horizon (the ``cosmological horizon'') such that the region of $\Sigma$ bounded by $B$ and $B'$ is compact. The canonical energy, $\ms E$, will get an additional contribution from $B'$, but this can be handled in the same manner as the contribution from $B$.} satisfying the vacuum Einstein equation with a cosmological constant \(\cc > 0\). Due to the cosmological constant in the Einstein equation, the ADM evolution equation \cref{eq:Ric-id} gets an additional term \(\tfrac{2}{d-1}N \cc h_{ab}\) on the right-hand-side, while, the constraint \cref{eq:H-constraint} gets an additional \(2\cc\) on the right-hand-side. Consequently, the right-hand-side of \cref{eq:useful} gets an extra term of \(\tfrac{2}{d-1}N \cc \delta^\Lambda{}_\Theta\). A repetition of the above calculation shows that we get the extra term $\tfrac{4}{d-1}\lambda \int_\Sigma 4 N (A_\Lambda)^2$ in \cref{finpos}. For asymptotically deSitter black holes with \(\cc > 0\), the additional term contributed by the cosmological constant is positive.
\end{remark}
\medskip

\begin{remark}[Extremal black hole backgrounds]\label{rem:ex-bh}
Our analysis also applies to axisymmetric electromagnetic perturbations with \emph{compact support} initial data on stationary-axisymmetric (asymptotically flat or deSitter) extremal black hole backgrounds. In particular, our results imply that the criterion for instability, in terms of violating the near-horizon ``effective Breitenl\"ohner-Freedman bound'', given in Sec.5.2 and 6.4 \cite{HI-ex-stab} is never satisfied for axisymmetric electromagnetic perturbations, in agreement with the numerical evidence in Sec.III.E \cite{DR}.
\end{remark}

\section{Other positive forms of $\ms U$ in $(3+1)$ and $(4+1)$ dimensions}\label{sec:U-pos-other}

In this section we will consider the special case of \(d=3\) with one axial Killing field (\(k=1\)), and \(d=4\) with two axial Killing fields (\(k=2\)). We further restrict to the case where, in differential forms notation,\footnote{Here \(*\) denotes the Hodge dual on \(\Sigma\) with respect to the metric \(h_{ab}\).} $*E$---which is closed, $d*E = 0$, by \cref{eq:gauss}---is exact, i.e., $*E = d \Gamma$ for some $(d-2)$-form $\Gamma$. By the deRham theorem, the necessary and sufficient condition for $*E$ to be exact is that the charge integral $\int_S *E$ vanish for all closed $(d-1)$-surfaces $S$. Let $S_\infty$ be a sphere near infinity. If $S_\infty$ generates the nontrivial $(d-1)$-homology of $\Sigma$, then $*E$ will be exact provided that the total charge, of the solution vanishes,
\be\label{eq:chargeless}
	\int_{S_\infty}  *E = \int_{S_\infty}  r^a E_a = 0
\ee
Note that in the axisymmetric case, \cref{eq:chargeless} only restricts the ``polar'' electric field \(\tilde E_a\).

The reason why the cases of \(d=3\) with one axial Killing field and \(d=4\) with two axial Killing fields are special is that in these cases, the manifold of orbits of $\phi_\Lambda^a$ is two dimensional. It follows that for axisymmetric polar electric fields, \(\tilde E_a\), the $(d-1)$-form $\Gamma$ can be replaced by a scalar potential $H$ via
\be\label{eq:E-potential}\begin{split}
	\tilde E_a & = \Phi^{-1/2} \tilde\varepsilon_a{}^{b} D_{b} H
\end{split}\ee
Here \(\tilde\varepsilon_{ab}\) is the volume form on the \(2\)-dimensional manifold of orbits of \(\phi^a_\Lambda\) given by
\be\begin{split}
	\tilde\varepsilon_{ab} \defn \frac{1}{k!} \Phi^{1/2}\varepsilon_{ab}{}^{c_1 \ldots c_k } \eta_{\Lambda_1 \ldots \Lambda_k} \phi_{c_1}^{\Lambda_1} \ldots \phi_{c_k}^{\Lambda_k}
\end{split}\ee
with \(\eta_{\Lambda_1 \ldots \Lambda_k}\) the \(k\)-form on \(\bb V\) defined below \cref{eq:det-Phi}. Note that \(\phi^a_\Lambda \tilde\varepsilon_{ab} = 0 \), and (using \cref{eq:Dphi}) we have \( D_{[c} (\Phi^{-1/2} \tilde\varepsilon_{ab]}) = 0 \).

Using the potential \(H\) we can integrate-by-parts the last two terms of \cref{eq:U-decomp} in another way as
\be\label{eq:IBP2}\begin{split}
	&\quad - \int_\Sigma 2 N^\Lambda \tilde E^a D_{a} A_\Lambda - \int_B 2 r^a \tilde E_a N^\Lambda A_\Lambda \\
	&= - \int_\Sigma 2 N^\Lambda D_b (\Phi^{-1/2}\tilde\varepsilon^{ab} H) D_{a} A_\Lambda - \int_B 2 r_a D_b (\Phi^{-1/2}\tilde\varepsilon^{ab} H) N^\Lambda A_\Lambda \\
	&= \int_\Sigma 2 (\Phi^{-1/2}\tilde\varepsilon^{ab} D_b N^\Lambda) H D_{a} A_\Lambda + \int_B 2 D_b N^\Lambda (r_a \Phi^{-1/2}\tilde\varepsilon^{ab} H A_\Lambda)
\end{split}\ee
The last term vanishes due to the background ADM evolution equation \cref{eq:Dshift}. Thus, by taking an arbitrary linear combination of \cref{eq:IBP1,eq:IBP2} we can write the potential energy as
\be\label{eq:U-H-A}\begin{split}
	4\pi\ms U &= \int_\Sigma N \Phi^{-1} (D_a H)^2 + N (D_a A_\Lambda)^2 \\
	&\qquad + 4N \Phi^{-1} \Phi^{1/2}\tilde\varepsilon^{ab} \pi_b^\Lambda \lb[q A_\Lambda D_a  H - (1-q)  H D_{a} A_\Lambda \rb]
\end{split}\ee
for any real constant \(q\). Defining
\be\label{eq:Y-choice}
	Y^{\Lambda a} \defn \Phi^{1/2}\tilde\varepsilon^{ab} \pi_b^\Lambda 
\ee
we can write \cref{eq:U-H-A} as 
\be\label{eq:exp1}\begin{split}
	4\pi \ms U &= \int_\Sigma N \Phi^{-1} \lb( D_a  H + 2q Y_{a}^\Lambda A_\Lambda  \rb)^2 + N (D_aA_\Lambda - 2(1-q) \Phi^{-1} Y_{\Lambda a}  H )^2 \\
	&\qquad  - 4N \Phi^{-1} q^2 (Y_{a}^\Lambda A_\Lambda)^2 - 4N \Phi^{-2}(1-q)^2 ( Y_{\Lambda a} H)^2
\end{split}\ee

For the first term in the last line of \cref{eq:exp1}, using \cref{eq:DX-id} and the computations in \cref{iden1,iden2} we get
\be\label{eq:id1}\begin{split}
	\int_\Sigma - 4N \Phi^{-1} (Y_{a}^\Lambda A_\Lambda)^2 & = \int_\Sigma - 4N \pi_{a}^\Lambda \pi^{a \Theta} A_\Lambda A_\Theta \\
	& = \int_\Sigma 4 N \lb[ X_a{}^{\Lambda \Theta} A_\Theta D^a A_\Lambda + X_a{}^{\Lambda \Xi} X^a{}^\Theta{}_\Xi A_\Lambda A_\Theta \rb]
\end{split}\ee

To similarly manipulate the last term in \cref{eq:exp1}, we define \(X_a = X_a{}^\Lambda{}_\Lambda\), which satisfies (from \cref{eq:DX-id})
\be\label{eq:DX-trace-id}
	D^a(N \Phi^{-1} X_a) = 2N \Phi^{-1} X_a^2 + 2 N \Phi^{-1} (\pi_{\Lambda a})^2
\ee
We obtain,
\be\label{eq:id2}\begin{split}
	\int_\Sigma - 4N \Phi^{-2} ( Y_{\Lambda a} H)^2 & = \int_\Sigma - 4N \Phi^{-1} ( \pi_{\Lambda a})^2 H^2 = \int_\Sigma 4N \Phi^{-1} \lb[ X^a H D_a H + X_a^2 H^2 \rb]
\end{split}\ee

Using \cref{eq:id1,eq:id2} the potential energy expression \cref{eq:exp1} can be written as 
\be\label{eq:exp2}\begin{split}
	4\pi \ms U &= \int_\Sigma N \Phi^{-1} \lb( D_a  H + 2q Y_{a}^\Lambda A_\Lambda  \rb)^2 + N (D_aA_\Lambda - 2(1-q) \Phi^{-1} Y_{\Lambda a}  H )^2 \\
	&\qquad + 4 N q^2 \lb[ X_a{}^{\Lambda \Theta} A_\Theta D^a A_\Lambda + X_a{}^{\Lambda \Xi} X^a{}^\Theta{}_\Xi A_\Lambda A_\Theta \rb] \\
	&\qquad + 4N \Phi^{-1}(1-q)^2 \lb[ X^a H D_a H + X_a^2 H^2 \rb] \\[1.5ex]
	&= \int_\Sigma N \Phi^{-1} \lb( D_a H + 2(1-q) X_a H + 2q Y_{a}^\Lambda A_\Lambda  \rb)^2 \\
	&\qquad + N (D_aA_\Lambda + 2 q X_a{}^\Theta{}_\Lambda A_\Theta - 2(1-q) \Phi^{-1} Y_{\Lambda a}  H )^2 \\
	&\qquad - 4 N q(1-q) \lb[ X_a{}^{\Lambda \Theta} A_\Theta D^a A_\Lambda + X^a H D_a H  \rb] \\
	&\qquad - 8 N \Phi^{-1} q(1-q) \lb[ X_a Y^{\Lambda a} - X_a{}^\Lambda{}_\Theta Y^\Theta{}^{a}\rb] A_\Lambda H
\end{split}\ee\\

First, consider the case \(d=4\) with two axial Killing fields. Since the last line of \cref{eq:exp2} is of indefinite sign we choose \(q=0\) or \(q=1\) giving us two positive forms of the potential energy
\begin{subequations}\label{eq:U-pos-4d-2k}\begin{align}
	\ms U & = \tfrac{1}{4\pi}\int_\Sigma N \Phi^{-1} \lb( D_{a}  H + 2 X_{a}  H \rb)^2 + N (D_aA_\Lambda - 2 \Phi^{-1} Y_{\Lambda a}  H )^2 \label{eq:U-pos-4d-2k-1}\\
	& = \tfrac{1}{4\pi}\int_\Sigma N \Phi^{-1} \lb( D_{a}  H + 2 Y_{a}^\Lambda A_\Lambda  \rb)^2 + N (D_aA_\Lambda + 2 X_a{}^\Theta{}_\Lambda A_\Theta )^2 \label{eq:U-pos-4d-2k-2}
\end{align}\end{subequations}
The second form \cref{eq:U-pos-4d-2k-2} (\(q = 1\)) is, of course, just the previously derived positive expression \cref{finpos} written in terms of the scalar potential \(H\). \\

Now, consider the case \(d=3\). Since there is only one axial Killing field we can drop the indices \(\Lambda, \Theta, \ldots\) Then, \(\Phi_{\Lambda \Theta} \equiv \Phi \) and the axial vector potential is of the form \(A \phi_a \) i.e. \(A_\Lambda \equiv \Phi A \). Similarly, \(Y^\Lambda_a \equiv Y_a = \varepsilon_a{}^{bc} \phi_c \pi_b\) and \(X_a{}^\Lambda{}_\Theta \equiv X_a = -\tfrac{1}{2}\Phi^{-1}D_a \Phi\). The last line of \cref{eq:exp2} vanishes identically and using \cref{eq:id1,eq:id2} in the second-to-last line of \cref{eq:exp2} we get the manifestly positive form
\be\label{eq:U-pos-3d}\begin{split}
	\ms U & = \tfrac{1}{4\pi}\int_\Sigma N \Phi^{-1} \lb( D_{a}  H + 2(1-q) X_{a}  H + 2q Y_{a} \Phi A  \rb)^2 \\
	&\qquad + N \Phi^{-1} \lb(D_a(\Phi A) + 2q X_a \Phi A - 2(1-q) Y_{a}  H \rb)^2 \\
	&\qquad + 4 N \Phi^{-1} q(1-q) \lb( X_a^2 + Y_a^2 \rb)  \lb(  H^2 + (\Phi A)^2 \rb)
\end{split}\ee
for any \(0 \leq q \leq 1\) with \cref{finpos} corresponding to \(q=1\). This expression for $q=1/2$ has been obtained independently by Gudapati \cite{G-em}.

The case $q=1/2$ for $d=3$ is particularly interesting. 
To see this, we replace $A$ and $H$ by a complex potential $\Psi$
\be\label{eq:Psi-defn}
\Psi \defn \Phi A - i H
\ee
Then with \(q = 1/2\) the potential energy \cref{eq:U-pos-3d} can be written as 
\be\label{eq:U-3d-Psi}
	\ms U = \tfrac{1}{4\pi}\int_\Sigma N \Phi^{-1} \bigg[ \abs{ D_a \Psi + Z_a \Psi }^2 + \abs{Z_a}^2 \abs{\Psi}^2 \bigg] 
\ee
where
\be\label{eq:Z-defn}
	Z_a \defn X_a + i Y_a = -\tfrac{1}{2}\Phi^{-1}D_a \Phi + i \varepsilon_a{}^{bc}\phi_c \pi_b \, .
\ee
Until this point, we have considered only the $t$-$\phi$ reflection odd data determined by $A$ and $H$. However, if $t$-$\phi$ even data also is present, the evolution equations \cref{eq:evol-EM} allow us to express this data in terms of $\dot{A}$ and $\dot{H}$ as
\be\begin{split}
	E_\Lambda \equiv \phi^a E_a & = N^{-1} \Phi\dot A \\
	D_{[a}\tilde A_{b]} & = -\tfrac{1}{2} N^{-1} \Phi^{-1} \varepsilon_{ab}{^c} \phi_c \dot H
\end{split}\ee
The kinetic energy \cref{eq:KE} is then given by
\be\label{eq:K-3d-Psi}
	\ms K = \tfrac{1}{4\pi}\int_\Sigma N^{-1} \Phi^{-1} \abs{\dot\Psi}^2
\ee

Next we show that the total canonical energy \(\ms E = \ms K + \ms U\) can be written in terms of an ``effective energy-momentum" tensor, ${\ms T}_{\mu\nu}$, for \(\Psi\). To do this we extend \(Z_a\) (\cref{eq:Z-defn}) to a complex \(1\)-form on spacetime \(Z_\mu\) so that \(t^\mu Z_\mu = \phi^\mu Z_\mu = 0\) and such that the pullback of \(Z_\mu\) to \(\Sigma\) is \(Z_a\). We use $Z_\mu$ to define a new complex derivative operator
\be\label{eq:new-D}
	\ms D_\mu \defn \nabla_\mu + Z_\mu \, .
\ee
We define
\be\label{eq:T-eff}
	\ms T_{\mu\nu}(\Psi) \defn \tfrac{1}{4\pi}\Phi^{-1} \bigg[ \bar{\ms D}_{(\mu} \bar\Psi \ms D_{\nu)} \Psi - \tfrac{1}{2} g_{\mu\nu}\lb( \abs{\ms D_\lambda \Psi}^2 + \abs{Z_\lambda}^2 \abs{\Psi}^2\rb) \bigg]
\ee
Using \cref{eq:U-3d-Psi,eq:K-3d-Psi} the canonical energy then can be written as 
\be
\ms E = \ms K + \ms U = 2\int_\Sigma {\ms T}_{\mu\nu} (\Psi)t^\mu u^\nu
\ee
Note that \({\ms T}_{\mu\nu}(\Psi)\) contains terms with no derivatives of \(\Psi\) while the Maxwell energy-momentum tensor \cref{eq:T-EM} depends only on the first derivatives of \(\Psi\).
Note also that the equations of motion \cref{eq:evol-EM}, when expressed in terms of \(\Psi\), take a ``wave-like'' form
\be\label{eq:Psi-eom}\begin{split}
	\ddot\Psi & = N (D^a + Z^a) \lb[ N(D_a + Z_a) \Psi \rb] - N^2 \abs{Z_a}^2 \Psi \\[1.5ex]
	\implies &\qquad  \ms D^\mu \ms D_\mu \Psi - \abs{Z_\mu}^2 \Psi = 0
\end{split}\ee
This ``wave-like" form \cref{eq:Psi-eom} along with the ``effective energy-momentum" \cref{eq:T-eff} may be useful for proving decay results for electromagnetic perturbations following \cite{FKSY, And, Tataru, DRS-stab}.

\begin{remark}[Relation to the self-dual Maxwell field strength]\label{rem:EM-sd}
	Consider the complex \emph{self-dual} part \(\ms F_{\mu\nu}\) of the Maxwell field tensor \(F_{\mu\nu}\)
	\be
		\ms F_{\mu\nu} \defn F_{\mu\nu} - i (* F)_{\mu\nu}
	\ee
	where \(*\) is the Hodge dual with respect to the background spacetime metric \(g_{\mu\nu}\). The reflection-even and -odd parts of \(\ms F_{\mu\nu}\) on \(\Sigma\) are given by
	\begin{subequations}\begin{align}
		\ms F^{\rm (even)}_{ab} & = - i N^{-1} \Phi^{-1} \varepsilon_{ab}{}^c \phi_c \dot\Psi \\
		\ms F^{\rm (odd)}_{ab} & = - 2\Phi^{-1} \phi_{[a} D_{b]} \Psi
	\end{align}\end{subequations}
	Thus, we can view \(\Psi\) and \(\dot\Psi\) as complex ``magnetic" potentials for \(\ms F_{\mu\nu}\). The corresponding ``electric" potentials can be obtained using the duality relation \((*\ms F)_{\mu\nu} = i \ms F_{\mu\nu}\).
\end{remark}
\medskip

\begin{remark}[Relation to the Ashtekar-Sen connection]\label{rem:Ash-Sen}
For the case \(d=3\), consider some set of frames \(e_a^i\) on \(\Sigma\) that are orthonormal with respect to the metric \(h_{ab}\) (here \(i,j,\ldots\) are abstract indices in \(\bb R^3\)). Let \(\omega_a{}^i{}_j\) be the corresponding \(\mf{so}(3)\)-valued spin connection. The \emph{anti-self-dual} \emph{Ashtekar-Sen connection} is given by
\be\begin{split}
	\mc A_a^i \defn \tfrac{1}{2}\epsilon^{ijk} \omega_{ajk} + i K_a{}^b e_b^i
\end{split}\ee
A direct computation shows that \(Z_a\) defined in \cref{eq:Z-defn} is 
\be
	Z_a = \Phi^{-1} \varepsilon_{abc} \phi^c \phi^d \mc A_d^i e^b_i 
\ee
Note that the \emph{anti-self-dual} Ashtekar-Sen connection comes along with the potential \(\Psi\) for the \emph{self-dual} electromagnetic field (see \cref{rem:EM-sd}) in \cref{eq:T-eff,eq:Psi-eom}.
\end{remark}
\medskip

\begin{remark}[Relation to the twist and Ernst potentials]\label{rem:twist-Ernst}
In \(d=3\) the constraint equation \cref{eq:diff-constraint} becomes \(D^a (\Phi \pi_a) = 0\), which (at least locally) can be solved using the \emph{twist potential} \(\omega\) \cite{Geroch,Mon-reduction} as
\be
	\pi_a  = \tfrac{1}{2}\Phi^{-2} \varepsilon_a{}^{bc}\phi_c D_b \omega
\ee
Defining a complex potential \(\Upsilon \defn \Phi + i \omega\), we can write \(Z_\mu\) as
\be
	Z_\mu = -\tfrac{1}{2} \Phi^{-1} \nabla_\mu \Upsilon
\ee
The complex potentials \(\Psi\) and \(\Upsilon\) are the \emph{Ernst potentials} for the electromagnetic perturbations and the vacuum gravitational background, respectively \cite{Ernst1, Ernst2}. It is known that the Einstein-Maxwell equations take the form of a wave map when written in terms of the Ernst potentials. This line of investigation has been pursued by Moncrief and Gudapati \cite{MG-stab}, who have succeeded in showing positivity of canonical energy for linearized Einstein-Maxwell perturbations on a Kerr-Newman background.
\end{remark}

\section*{Acknowledgements}

K.P. is supported in part by the NSF grants PHY-1404105 and PHY-1707800 to Cornell University. R.M.W. is supported in part by NSF grant PHY~15-05124 to the University of Chicago.

\appendix

\section{Non-positivity of $\ms E_F$}\label{sec:non-positivity}

In this appendix, we will show that for any rotating black hole background spacetime that satisfies the vacuum Einstein equation, there always exist axisymmetric electromagnetic perturbations that make $\ms E_F$ negative, where $\ms E_F$ is the energy, \cref{eq:E-T1}, obtained from the Maxwell stress-energy tensor. This shows that it is essential to use $\ms E$ rather than $\ms E_F$ in stability arguments.
Our proof is based upon the positive form \cref{finpos}, that we obtained for $\ms E$ for $t$-$\phi$-reflection-odd electromagnetic fields together with the fact that 
\be
	2\ms E_F =  \ms E+ \tfrac{1}{4\pi} \int_B 2 r^a E_a N^b A_b 
\ee
(see \cref{eet}).

We first introduce local coordinates on a suitable region of $B$ as follows: The axial Killing fields $\phi_\Lambda^a$ are tangent to $B$ and are surface orthogonal on $B$. Let $\ms S$ denote such an orthogonal surface, and let $V \subset \ms S$ denote an open neighborhood in $\ms S$ that does not contain any axis points and can be covered by a coordinate patch $(x^1, \dots, x^{d-k-1})$, where $k$ is the number of axial Killing fields. Note that since there can be at most $d/2$ commuting axial Killing fields in an asymptotically flat $(d+1)$-dimensional spacetime, for $d \geq 3$ the orthogonal hypersurface must be at least $1$-dimensional, so there always must be at least one $x$-coordinate. Let $V_\phi$ denote the orbit of $V$ under the flow of the axial Killing fields $\phi_\Lambda^a$. Then $V_\phi$ is an axisymmetric open region of $B$ that contains no axis points and can be covered by coordinates $(x^1, \dots, x^{d-k-1}, \phi^1, \dots, \phi^k)$ such that in these coordinates, $\phi_\Lambda^a = (\partial/\partial \phi^\Lambda)^a$. Let $F(x^1,\dots, x^{d-k-1})$ be a smooth, non-negative (and not identically vanishing) function of compact support in $V_\phi$.

Now extend the coordinates $(x^1, \ldots, x^{d-k-1}, \phi^1, \ldots, \phi^k)$ to Gaussian normal coordinates \((r, x^1, \ldots, x^{d-k-1}, \phi^1, \ldots, \phi^k)\) covering a region $W$ of $\Sigma$, with $r=0$ corresponding to $V_\phi$. Since $B$ is compact, there is a $c > 0$ such that these coordinates are well defined for all $r \leq c$. Let $f: [0, \infty) \to {\bb R}$ be a smooth, nonnegative function with support in $[0, c)$ such that $f(0) = 1$. Let $\beta > 1$. We now make the following choices for $A_a$ and $E^a$:
\begin{subequations}\begin{align}
	A_a & = N^1 \sin (\beta x^1) f(\beta r) F(x^1,\dots, x^{d-k-1}) (d \phi^1)_a \\
	E^a & = D_b U^{ab}
\end{align}\end{subequations}
with
\be
U^{ab} = \cos(\beta x^1) f(\beta r) F(x^1,\dots, x^{d-k-1}) \left(\frac{\partial}{\partial r} \right)^{[a}
\left(\frac{\partial}{\partial x^1} \right)^{b]} \, .
\ee
Then $A_a$ and $E^a$ are axisymmetric, and $D_a E^a = 0$. They also are $t$-$\phi$-reflection-odd, so $\ms K = 0$ and $\ms E = \ms U$.

Now, on $B$, for large $\beta$, we have
\be
r^a E_a = - \beta \sin(\beta x^1) F(x^1,\dots, x^{d-k-1}) + O(1)
\ee
where by ``$O(1)$'' we mean a function that is uniformly bounded as $\beta \to \infty$. Thus, on $B$ we have
\be
r^a E_a N^b A_b = - \beta (N^1)^2 \sin^2(\beta x^1) F^2(x^1,\dots, x^{d-k-1}) + O(1) \, .
\label{bndy}
\ee
It follows that for sufficiently large $\beta$, we have
\be\label{eq:bndy-int}
 \int_B 2 r^a E_a N^b A_b < - C \beta
\ee
for some $C > 0$.

On the other hand, by inspection of the explicit expressions for $A_a$ and $E_a$, for large $\beta$ we clearly have
\be
|A_a| \leq c_1 \eqsp |D_a A_b| \leq c_2 \beta \eqsp |E_a| \leq c_3 \beta
\ee
everywhere on $\Sigma$, for some constants $c_1, c_2, c_3$. Furthermore, the lapse function $N$ is bounded by
\be
N \leq c_4 r
\ee
Since $A_a$ and $E_a$ are nonvanishing only for $r < c/\beta$, it follows immediately from our expression \cref{finpos} for $\ms E = \ms U$ that for large $\beta$
\be
|\ms E| \leq c_5 \beta^2 \int_0^{c/\beta} r dr \leq C'
\ee
for some constant $C'$. Thus, for sufficiently large $\beta$, the negative boundary term \cref{eq:bndy-int} dominates over $\ms E$, thereby making $\ms E_F < 0$.




\bibliographystyle{JHEP}
\bibliography{em-energy}       

\providecommand{\href}[2]{#2}\begingroup\raggedright\begin{thebibliography}{10}

\bibitem{Ch-Kl}
D.~Christodoulou and S.~Klainerman, {\em {The global nonlinear stability of the
  Minkowski space}}.
\newblock Princeton University Press, 1993.

\bibitem{RW}
T.~Regge and J.~A. Wheeler, {\it {Stability of a Schwarzschild singularity}},
  {\em Phys. Rev.} {\bf 108} (1957) 1063--1069.

\bibitem{Zerilli}
F.~J. Zerilli, {\it {Effective potential for even parity Regge-Wheeler
  gravitational perturbation equations}},  {\em Phys. Rev. Lett.} {\bf 24}
  (1970) 737--738.

\bibitem{IK-stab}
A.~Ishibashi and H.~Kodama, {\it {Stability of higher dimensional Schwarzschild
  black holes}},  {\em Prog. Theor. Phys.} {\bf 110} (2003) 901--919,
  [\href{http://arxiv.org/abs/hep-th/0305185}{{\tt hep-th/0305185}}].

\bibitem{DHR}
M.~Dafermos, G.~Holzegel, and I.~Rodnianski, {\it {The linear stability of the
  Schwarzschild solution to gravitational perturbations}},
  \href{http://arxiv.org/abs/1601.06467}{{\tt arXiv:1601.06467}}.

\bibitem{ER}
R.~Emparan and H.~S. Reall, {\it {Black Holes in Higher Dimensions}},  {\em
  Living Rev. Rel.} {\bf 11} (2008) 6,
  [\href{http://arxiv.org/abs/0801.3471}{{\tt arXiv:0801.3471}}].

\bibitem{FKSY}
F.~Finster, N.~Kamran, J.~Smoller, and S.-T. Yau, {\it {Decay of Solutions of
  the Wave Equation in the Kerr Geometry}},  {\em Commun. Math. Phys.} {\bf
  264} (2006), no.~2 465--503.

\bibitem{And}
L.~Andersson and P.~Blue, {\it {Hidden symmetries and decay for the wave
  equation on the Kerr spacetime}},  {\em Ann. of Math. (2)} {\bf 182} (2015),
  no.~3 787--853, [\href{http://arxiv.org/abs/0908.2265}{{\tt
  arXiv:0908.2265}}].

\bibitem{Tataru}
D.~Tataru, {\it {Local decay of waves on asymptotically flat stationary
  space-times}},  {\em {Am. J. Math.}} {\bf 135} (2013), no.~2 361--401,
  [\href{http://arxiv.org/abs/0910.5290}{{\tt arXiv:0910.5290}}].

\bibitem{DRS-stab}
M.~Dafermos, I.~Rodnianski, and Y.~Shlapentokh-Rothman, {\it {Decay for
  solutions of the wave equation on Kerr exterior spacetimes III: The full
  subextremal case $|a|< M$}},  {\em Ann. of Math. (2)} {\bf 183} (2016), no.~3
  797--913, [\href{http://arxiv.org/abs/1402.7034}{{\tt arXiv:1402.7034}}].

\bibitem{Blue}
P.~{Blue}, {\it {Decay of the Maxwell field on the Schwarzschild manifold}},
  {\em J. Hyperbolic Differ. Equ.} {\bf 05} (2008), no.~04 807--856,
  [\href{http://arxiv.org/abs/0710.4102}{{\tt arXiv:0710.4102}}].

\bibitem{ABB}
L.~Andersson, T.~Bäckdahl, and P.~Blue, {\it {Decay of solutions to the
  Maxwell equation on the Schwarzschild background}},  {\em Class. Quant.
  Grav.} {\bf 33} (2016), no.~8 085010,
  [\href{http://arxiv.org/abs/1501.04641}{{\tt arXiv:1501.04641}}].

\bibitem{Pasq}
F.~Pasqualotto, {\it {The spin $\pm$1 Teukolsky equations and the Maxwell
  system on Schwarzschild}},  \href{http://arxiv.org/abs/1612.07244}{{\tt
  arXiv:1612.07244}}.

\bibitem{AB}
L.~Andersson and P.~Blue, {\it {Uniform energy bound and asymptotics for the
  Maxwell field on a slowly rotating Kerr black hole exterior}},  {\em J.
  Hyperbolic Differ. Equ.} {\bf 12} (2015), no.~04 689--743,
  [\href{http://arxiv.org/abs/1310.2664}{{\tt arXiv:1310.2664}}].

\bibitem{SW-tphi}
J.~S. Schiffrin and R.~M. Wald, {\it {Reflection Symmetry in Higher Dimensional
  Black Hole Spacetimes}},  {\em Class. Quant. Grav.} {\bf 32} (2015), no.~10
  105005, [\href{http://arxiv.org/abs/1501.02752}{{\tt arXiv:1501.02752}}].

\bibitem{HW-stab}
S.~Hollands and R.~M. Wald, {\it {Stability of black holes and black branes}},
  {\em Commun. Math. Phys.} {\bf 321} (2013) 629--680,
  [\href{http://arxiv.org/abs/1201.0463}{{\tt arXiv:1201.0463}}].

\bibitem{PW}
K.~Prabhu and R.~M. Wald, {\it {Black Hole Instabilities and Exponential
  Growth}},  {\em Commun. Math. Phys.} {\bf 340} (2015), no.~1 253--290,
  [\href{http://arxiv.org/abs/1501.02522}{{\tt arXiv:1501.02522}}].

\bibitem{FMR}
P.~Figueras, K.~Murata, and H.~S. Reall, {\it {Black hole instabilities and
  local Penrose inequalities}},  {\em Class. Quant. Grav.} {\bf 28} (2011)
  225030, [\href{http://arxiv.org/abs/1107.5785}{{\tt arXiv:1107.5785}}].

\bibitem{SW-turning-pt}
J.~S. Schiffrin and R.~M. Wald, {\it {Turning Point Instabilities for
  Relativistic Stars and Black Holes}},  {\em Class. Quant. Grav.} {\bf 31}
  (2014) 035024, [\href{http://arxiv.org/abs/1310.5117}{{\tt
  arXiv:1310.5117}}].

\bibitem{KW-stab}
B.~S. Kay and R.~M. Wald, {\it {Linear stability of Schwarzschild under
  perturbations which are nonvanishing on the bifurcation two-sphere}},  {\em
  Class. Quant. Grav.} {\bf 4} (1987) 893--898.

\bibitem{Daf-Rod-lec}
M.~Dafermos and I.~Rodnianski, {\it {Lectures on black holes and linear
  waves}},  {\em Clay Mathematics Proceedings} (17) 97--205,
  [\href{http://arxiv.org/abs/0811.0354}{{\tt arXiv:0811.0354}}].

\bibitem{Wald-book}
R.~M. Wald, {\em {General Relativity}}.
\newblock The University of Chicago Press, 1984.

\bibitem{CB-book}
Y.~Choquet-Bruhat, {\em {General Relativity and the Einstein Equations}}.
\newblock Oxford University Press, 2009.

\bibitem{RT}
T.~Regge and C.~Teitelboim, {\it {Improved Hamiltonian for general
  relativity}},  {\em Phys. Lett.} {\bf 53B} (1974) 101--105.

\bibitem{HI-ex-stab}
S.~Hollands and A.~Ishibashi, {\it {Instabilities of extremal rotating black
  holes in higher dimensions}},  {\em Commun. Math. Phys.} {\bf 339} (2015),
  no.~3 949--1002, [\href{http://arxiv.org/abs/1408.0801}{{\tt
  arXiv:1408.0801}}].

\bibitem{DR}
M.~Durkee and H.~S. Reall, {\it {Perturbations of near-horizon geometries and
  instabilities of Myers-Perry black holes}},  {\em Phys. Rev.} {\bf D83}
  (2011) 104044, [\href{http://arxiv.org/abs/1012.4805}{{\tt
  arXiv:1012.4805}}].

\bibitem{G-em}
N.~Gudapati, {\it {A Positive-Definite Energy Functional for Axially Symmetric
  Maxwell's Equations on Kerr-de Sitter Black Hole Spacetimes}},
  \href{http://arxiv.org/abs/1710.11294}{{\tt arXiv:1710.11294}}.

\bibitem{Geroch}
R.~P. Geroch, {\it {A method for generating solutions of Einstein's
  equations}},  {\em J. Math. Phys.} {\bf 12} (1971) 918--924.

\bibitem{Mon-reduction}
V.~Moncrief, {\it {Reduction of Einstein's equations for vacuum space-times
  with spacelike U(1) isometry groups}},  {\em Annals of Physics} {\bf 167}
  (1986), no.~1 118--142.

\bibitem{Ernst1}
F.~J. Ernst, {\it {New Formulation of the Axially Symmetric Gravitational Field
  Problem}},  {\em Phys. Rev.} {\bf 167} (1968) 1175--1179.

\bibitem{Ernst2}
F.~J. Ernst, {\it {New Formulation of the Axially Symmetric Gravitational Field
  Problem. II}},  {\em Phys. Rev.} {\bf 168} (1968) 1415--1417.

\bibitem{MG-stab}
V.~Moncrief and N.~Gudapati. in preparation.

\end{thebibliography}\endgroup
\end{document}